\documentstyle[12pt,graphicx]{article}
\newcommand{\be}{\begin{equation}}
\newcommand{\ee}{\end{equation}}
\newcommand{\ba}{\begin{eqnarray}}
\newcommand{\ea}{\end{eqnarray}}
\newcommand{\baa}{\begin{eqnarray}}
\newcommand{\eaa}{\end{eqnarray}}
\newcommand{\ed}{\end{document}}
\newcommand{\lab}[1]{\label{#1}}
\newcommand{\re}[1]{(\ref{#1})}
\newcommand{\ci}[1]{\cite{#1}}
\renewcommand{\baselinestretch}{1}
\date{}
\begin{document}
\title{Casimir Effect for spherical boundaries: A thermofield dynamics approach}
\author{D.U.~Matrasulov\thanks{Heat Physics Department of the Uzbek Academy of Sciences,
28 Katartal St., 700135 Tashkent, Uzbekistan}, Kh.T.~Butanov$^*$, Kh.Yu.~Rakhimov$^*$\\
F.C.~Khanna\thanks{Physics Department University of Alberta Edmonton Alberta, T6G 2J1 Canada and TRIUMF, 4004 Wersbrook Mall, Vancouver, British Columbia, Canada, V6T2A3},
and A.E.~Santana\thanks{Instituto de F\'isica, Universidade de Bras\'ilia, 70910-900, Bras\'ilia-DF, Brazil}}
\maketitle

\begin{abstract}

The Casimir effect for spherical geometry is calculated using generalized
Thermofield Dynamics for the case of scalar field. Casimir force and Casimir
pressure are presented. It is found that for high temperatures the Casimir
force does change sign.

\end{abstract}


\section{Introduction}

Casimir effect and its applications have been a topic of extensive
experimental and theoretical studies since its first prediction in 1948 \ci{1}-\ci{steve}. It
has been measured in experiments \ci{2}-\ci{most2} involving spherical metallic and non-mettalic balls \ci{most2} and has
thus led to its application as switching mechanism in nano-technology. Role
of finite temperature on the Casimir effect has been considered theoretically \ci{most2}-\ci{milton1}
but, in general, the magnitude of the force is too small to be measured
precisely. Recently Casimir-Polder force, along with the effect of finite
temperature, has been measured between a bulk object and a gas-phase atom \ci{obr}.
Effect of finite temperature on Casimir effect, force between two bulk
such as two metallic objects or two dielectric objects, has eluded
detection due to the fact that the variation with temperature is very small.
However it is becoming important to calculate the Casimir force in a
spherical geometry that involves particles. Such is the case of baryons, in
particle physics, that have quark confinement and it is expected that the
quarks deconfine at high temperature. This is the central objective at
facilities such as RHIC, that allows collisions between heavy ion beams, and
LHC, that will allow collisions between heavy ion beams like Pb at much
higher energies. The object is to de-confine the quarks and thus form a quark-gluon
plasma, a form of matter that is believed to have existed soon after the big
bang. The cosmology suggests that, the free quarks and gluons, as they cool down, combine
into a confined state that are the baryons that eventually lead to the
formation of nuclei at a later epoch of the universe. This suggests that
the Casimir energy, in particular at finite temperature, would play an important
role in the confinement and then the de-confinement of quarks and gluons. However we
know that the dynamics of quarks and gluons is based on a non-Abelian theory
of strong interactions, Quantum Chromo-Dynamics(QCD). Such a theory presents
its own sort of difficulties due its non-Abelian nature. In this paper we
will study the case of spherical geometry of the confined system but with
an Abelian theory, in particular, for a scalar field. Studies with QCD will be
considered separately. The present
study uses the techniques developed recently and are based on using
thermofield dynamics (TFD), real time finite temperature field theory, to get
the finite temperature effects. This
approach depends on using the Bogoliubov transformation to include the
temperature in quantum field theory \ci{ademir,ademir1,ndfi04}. It is well-known that finite
temperature Green's functions obey KMS condition that obeys periodicity (for
Bosons) or anti-periodicity (Fermions) condition in temperature. This may be
viewed as a confinement condition with temperature \ci{um}-\ci{das}. This has been extended
with a great deal of success to the study of Casimir effect in a box or a
paralellopiped. This technique will be used first for zero temperature case
and then the spherical geometry. Then the Casimir effect may be interpreted
as a vacuum condensation of the field in a confined geometry. Such an
interpretation follows the original work that gave an elegant understanding
of Superconductivity. The study will focus on calculating the Casimir energy
and Casimir force for a field confined in a spherical geometry when the role
of finite varying temperature is included. The study is based on the
earlier work of Bender and Hays~\ci{bender} to finite temperature using the techniques
mentioned above. Additional work at zero temperature has been carried by
several authors \ci{boyer}-\ci{hagen2}. It is important to note that our technique will allow a
variation with size of the spherical bag and with temperature quite easy. In section~\ref{tfd},
necessary details of TFD
will be presented. Section~\ref{cesb} gives details of the Casimir effect at $T=0$. In
section \ref{ft}, application of TFD to the calculation of the Casimir effect at
finite temperature will be provided. In section \ref{rd}, numerical results and
their comparison to earlier calculations will be shown. Variations with
$T$ and size of the sphere will be displayed. Finally some concluding
remarks and future directions will be presented.

\section{Generalization of TFD}\label{tfd}

The idea of compactification is based on the Bogoliubov transformations (BT).
Briefly the idea is the following:
The Green's function according to TFD prescription is given by a $2\times 2$ matrix
with the elements
\begin{equation}
G^{11} = G_0 +v^2(\omega,\beta)[G_0 -G^*_0],
\end{equation}
$$
G^{22} = -G_0^* +v^2(\omega,\beta)[G_0 -G_0^*],
$$
$$
G^{12} = G^{21}= v(\omega,\beta)[1+v^2(\omega,\beta)]^{1/2}[G_0 -G_0^*]
$$
with
$$
v^2(\omega,\beta) = \frac{1}{e^{\beta \mid\omega\mid}-1}.
$$
The physically observable quantities are related to $G^{11}$ only \ci{ademir}.

The finite-temperature Green's function in TFD is expressed in terms of
the zero-temperature one using the Bogoliubov transformations
which are defined as (for the case of bosons)
\begin{equation}
v^2(\omega,\beta) = \frac{1}{e^{\beta\omega} -1} = \sum\limits_{l=1}^{\infty} e^{-l\beta\omega}.
\end{equation}

The finite-temperature Green's function in the momentum representation is expressed via the zero-temperature one as follows
(here we write only $11-$element of the $2\times 2$-matrix, while other elements can be obtained similarly):
\begin{equation}
G^{k,\beta} = G_0(k) +v^2(k,\beta)[G_0(k) +\tilde G_0(k)]
\end{equation}
which defines the Green's function in the coordinate representation as
\begin{equation}
G^{11}(x-x',\beta) = \frac{1}{(2\pi)^2}\int d^4q\, G^{11}(q,\beta) e^{iq(x-x')} = G^{11}({\vec x}-{\vec x'}, x_0-i\beta)
\end{equation}
where $q$ and $x$ are four-vectors.

These transformations can be generalized by introducing the factor
\begin{equation}
v^2(q, \alpha) = \sum\limits_{l=1}^{\infty} e^{-i\alpha_l q}
\end{equation}
where the notation
\begin{equation}
\sum\limits_{l=1}^{\infty} e^{-i\alpha_l \cdot q} = \sum\limits_{l_0,l_1,l_2,l_3=1}^{\infty} exp[-i(\alpha_0l_0q_0 +\alpha_1l_1q_1 +\alpha_2l_2q_2 +\alpha_3l_3q_3)]
\end{equation}
is used.

Then the Green's function obtained by applying the Bogoliubov transformation
is written as
\begin{eqnarray}
&G^{11}(x-x',\alpha) = \frac{1}{(2\pi)^4}\int d^4q\, G^{11}(q,\alpha) e^{iq(x-x')} = \nonumber \\
&\frac{1}{(2\pi)^4}\int d^4q \, e^{iq(x-x')} \sum\limits_{l=1}^{\infty}e^{-i\alpha_l \cdot q} [G^{11}_0(q) +\tilde G^{11}_0(q)] = \nonumber \\
&2\sum\limits_{l=1}^{\infty} G^{11}(x-x'-\alpha_l).
\end{eqnarray}
Thus the finite-temperature Green's function within the generalized TFD prescription
is obtained from the zero-temperature Green's function using the Bogoliubov transformation.
In section~\ref{ft} we will use this Green's function to calculate the Casimir energy at finite temperature.

\section{Casimir energy for spherical boundaries}\label{cesb}

The Casimir energy for spherical boundaries
is explored in several approaches and by many authors (for review see \ci{mull}).
In this section we give a brief description of the derivation of the Casimir energy based on the Green's function approach
following Bender and Hays~\ci{bender}.
In the Green's function based approach the Casimir energy is calculated as
\ci{mull,bender}
\begin{equation}
E_{C}= \lim\limits_{\tau \to 0}\frac{\partial^2}{\partial\tau^2}
\int d^3x\, \Gamma({\vec x},{\vec x},\tau)
\lab{eqq}
\end{equation}
where $\Gamma({\vec x},{\vec x},\tau)$ is the inhomogeneous part of the Green's
function which is given by
$$
G({\vec x},{\vec x'}, x-x_0) = G_0(x-x') + \Gamma({\vec x},{\vec x},x-x_0).
$$
For the spherical boundaries the (Fourier transformed) Green's function is given
by (for Dirichlet boundary conditions)
\begin{equation}
G({\vec x},{\vec x'}, \omega) = ik\sum\limits_{l=0}^{\infty}\left[j_l(kr')h^{(1)}_{l}(kr)-
\frac{h^{(1)}_{l}(kR)}{j_l(kR)}j_l(kr')j_l(kr)\right]\sum\limits_{m=-l}^{l} Y_{lm}(\Omega)
Y_{lm}(\Omega')
\lab{gr1}
\end{equation}
where $h^{(1)}_{l}(x)$ is the Hankel function of the first order, $j_l(kr)$ is
the spherical Bessel function, $k =\mid\omega\mid$ and $R$ is radius of the sphere.
Second term in this expression is the inhomogeneous part of the Green's function:
\begin{equation}
\Gamma({\vec x},{\vec x'}, \omega) = -ik\sum\limits_{l=0}^{\infty}\left[
\frac{h^{(1)}_{l}(kR)}{j_l(kR)}j_l(kr')j_l(kr)\right]
\sum\limits_{m=-l}^{l} Y_{lm}(\Omega) Y_{lm}(\Omega').
\lab{eq}
\end{equation}
Then the Casimir energy is written as
\begin{equation}
E(R,\tau) =\sum\limits_{l=0}^{\infty}(2l+1)\frac{\partial^2}{\partial \tau^2}\int\limits_{-\infty}^{\infty}\frac{d\omega}{2\pi}\,
e^{-i\omega \tau} k\frac{h^{(1)}_{l}(kR)}{j_l(kR)}\int\limits_{0}^{R}dr\, r^2j_l^2(kr)
\lab{eq00}
\end{equation}
where the orthogonality relation
$$
\sum\limits_{m=-l}^{l} Y_{lm}^2(\Omega) =\frac{2l+1}{4\pi}
$$
is used.

The radial integral can be expressed in terms of Bessel functions:
$$
\int\limits_{0}^{R}dr\, r^2j_l^2(kr) =\frac{R^2}{2}[j_{l}^2(kR)-
j_{l-1}(kR)j_{l+1}(kR)]$$
which allows us to write eq.~\re{eq00} as
\begin{equation}
E(R,\tau) =-\frac{1}{2}\sum_{l=0}^{\infty}(2l+1)\int\limits_{-\infty}^{\infty}\frac{d\omega}{2\pi}\,
e^{-i\omega \tau}i(kR)^3\frac{h^{(1)}_{l}(kR)}{j_l(kR)}[j_{l}^2(kR)-
j_{l-1}(kR)j_{l+1}(kR)].\lab{eq01}
\lab{eqn}
\end{equation}
Using the Wick rotations
$$
\omega \to i\omega, \;\;\; \tau \to i\tau, \;\;\; k \to i\mid\omega\mid
$$
and the substitutions
$$
x =\mid\omega\mid R, \;\;\; \delta = \tau/R
$$
this expression is written in the form
$$
E(R,\tau) =\lim\limits_{\delta\to 0}\frac{1}{2\pi R}
\sum_{l=0}^{\infty}(2l+1)\int\limits_{0}^{\infty}dx\, x^2\cos (\delta x)
\frac{K_{l+1/2}(x)}{I_{l+1/2}(x)}[I_{l+1/2}^2(x)-I_{l-1/2}(x)I_{l+3/2}(x)].
$$

In this expression the integral does not contain $R-$ dependance. Therefore the Casimir energy is
proportional to $R^{-1}$.

\section{Calculation at finite temperature}\label{ft}

Now we consider the Casimir effect at finite temperature. The generalised TFD
formalism is used when the field is coupled to a heat bath.
The above discussed generalized thermofield dynamics formalism can be used in this case to calculate the zero-point energy
at finite temperature.
To do this we need to rewrite the generalized BT, which can be written in spherical coordinates.

Then the Green's function is written as
\begin{equation}
G({\vec x},{\vec x'}, x-x_0) = \frac{1}{(2\pi)^4}\int d^3q\,dq_0\, G({\vec q},q_0) e^{i{\vec q}({\vec r}-{\vec r'})}e^{i q_0x_0}.
\end{equation}
With the prescription for the generalized BT, $G({\vec q},q_0)$ is written as
\begin{equation}
G({\vec q}, q_0,{\vec \alpha}, \alpha_0) = \sum\limits_{\vec{l},l_{0}=1}^{\infty}e^{-i{\vec \alpha_{\vec l}} \cdot {\vec q}}e^{-iq_{0}\alpha_{l_0}}G({\vec q},q_0)
\end{equation}
where $\vec{l}=(l_{r}, l_{\theta}, l_{\phi})$, $\vec{\alpha}_{{\vec l}}=(l_{r}\alpha_{r}, l_{\theta}\alpha_{\theta}, l_{\phi}\alpha_{\phi})$ and $\alpha_{l_0}=l_0\alpha_0$
\begin{eqnarray}
&G({\vec x},{\vec x'}, x-x_0, {\vec \alpha},\alpha_0) = \frac{1}{(2\pi)^4}\int d^3q\,dq_0\, e^{i{\vec q}({\vec r}-{\vec r'})} e^{iq_0x_0}
\sum\limits_{\vec{l},l_0=1}^{\infty}e^{-i{\vec \alpha_{\vec l}} \cdot {\vec q}} e^{i\alpha_{l_0} q_0} G_0({\vec q}, q_0) = \nonumber \\
&\frac{1}{(2\pi)^4}\sum\limits_{\vec{l},l_0=1}^{\infty}\int d^3q\,dq_0\, e^{i{\vec q}({\vec r}-{\vec r'} - {\vec \alpha_{\vec l}})}
e^{iq_0(x_0-\alpha_{l_0})} G_0({\vec q}, q_0).
\end{eqnarray}
Denoting ${\vec r''} = {\vec r'} -{\vec \alpha_{\vec l}}$
we have
\begin{equation}
G({\vec x},{\vec x'}, x-x_0, {\vec \alpha}, \alpha_0) =
\frac{1}{(2\pi)^4}\sum\limits_{\vec{l},l_0=1}^{\infty}\int d^3q\,dq_0\, e^{i{\vec q}({\vec r}-{\vec r''})}
e^{iq_0(x_0-\alpha_{l_0})} G_0({\vec q}, q_0) \equiv G({\vec r},{\vec r''}, x_0-\alpha_0)
\end{equation}

Then for the compactified finite-temperature Green's function for spherical boundries (for Boson case) is given as
\begin{equation}
G({\vec x},{\vec x'}, \alpha) = ik_1\sum\limits_{l=0}^{\infty}\sum\limits_{\vec{l},l_0=1}^{\infty}[j_l(k_1r'')h^{1}_{l}(k_1r)-
\frac{h^{1}_{l}(k_1R)}{j_l(k_1R)}j_l(k_1r'')j_l(k_1r)]\sum\limits_{m=-l}^{l} Y_{lm}(\Omega')
Y_{lm}(\Omega'')
\lab{gr2}
\end{equation}

where, $k_1 = \mid\omega -il_0\alpha_0\mid$, $\Omega'$ is the angle between $r$ and $r''$, $r'' =r'-l_r\alpha_r$.
For the compactified Casimir energy we have from eq. \re{eq01}
\begin{equation}
E(R,\alpha) =\lim\limits_{\tau \to -il_0\alpha_0}\sum\limits_{l_0,l_r=1}^{\infty}\sum\limits_{l=0}^{\infty}(2l+1)\frac{\partial^2}{\partial \tau^2}\int\limits_{-\infty}^{\infty}\frac{d\omega}{2\pi}\,
e^{-i(\omega-il_0\alpha_0) \tau} k_1\frac{h^{1}_{l}(k_1R)}{j_l(k_1R)}
L(R,k_1,\alpha_r)
\lab{eq02}
\end{equation}
where
\begin{equation}
L(R,k_1,\alpha_r)\equiv \int\limits_{0}^{R}dr\, r^2j_l(k_1r)j_l(k_1(r-l_r\alpha_r)).
\end{equation}

Unlike the case of zero-temperature, when $\alpha_0=\alpha_r =0$,
$R$-dependence is not factorized in the case of finite temperature and thus this dependence cannot be written
explicitly.

\section{Results}\label{rd}
\begin{figure}[htbp]
\begin{center}
\includegraphics[width=9cm, height=6.5cm]{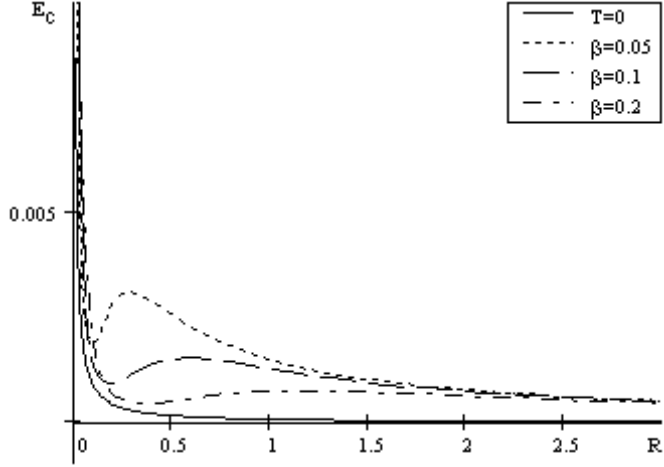}
\end{center}
\caption{Dependences of the Casimir energy on the sphere radius at
different temperatures and fixed $\alpha_r$ ($\alpha_r =10^{-6}$)
are compared with the corresponding zero-temperature result plotted
from ref.\ci{bender}.} \label{fig1}
\end{figure}

The Casimir energy is calculated numerically for a sphere with radius
$a$ for several values of $\beta$ and the compactification
parameter, $\alpha_r$. In the case of zero temperature
calculations of the Casimir energy diverge due to the divergence
of the integral for the Green's function over $r$
(between limits from $0$ to $a$). This divergence can be
regularized by replacing upper integration limit by
$a(1-\epsilon)$ with $epsilon \to 1$. In the case of
finite-temperature such a divergence
is suppressed due to the presence of the factor $e^{-l_0\alpha_0\tau}$
in eq.~\re{eq02}. In Fig.~\ref{fig1} Casimir energy is plotted as a function of the
radius for different temperatures at the fixed value of
$\alpha_r =10^{-6}$ and compared with the Casimir energy at zero
temperature.

The lowest curve in this figure shows the Casimir energy for
$T=0$ calculated on the basis of the approach developed by
Bender and Hays (using $\epsilon$-type cutoff). The following
values of $\beta$ are chosen: $\beta =0.2$, $\beta =0.1$ and
$\beta =0.05$.
It is clear that, in certain interval of $R$, an increase of the temperature
leads to considerable changes in
the $E(R)$ and for higher enough values of $T$ this curve has a
maximum, which corresponds to a sign change of the Casimir
pressure(derivative of the $E(R)$ over with $R$).
\begin{figure}[htbp]
\begin{center}
\includegraphics[width=15cm, height=6cm]{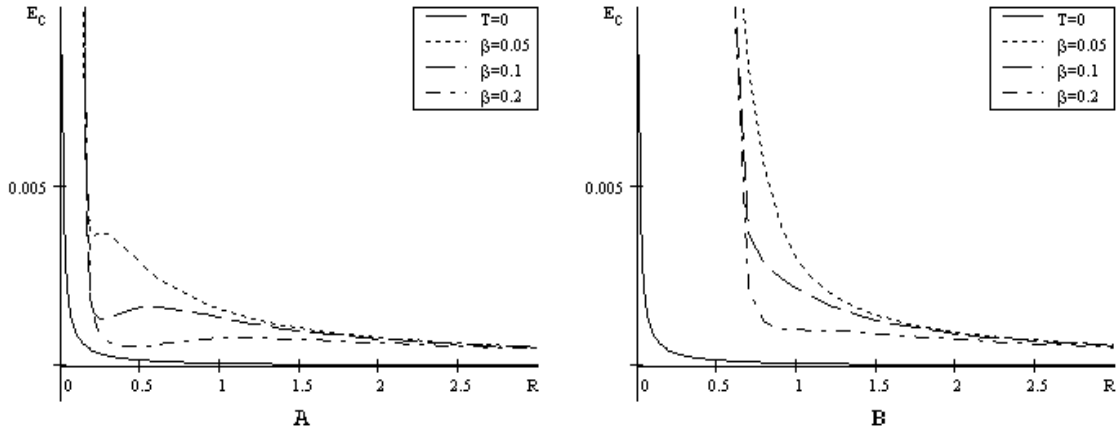}
\end{center}
\caption{Dependence of the Casimir energy at different
temperatures and for fixed $\alpha_r$. A: for $\alpha_r =0.1$; B:
$\alpha_r =0.5$.} \label{fig2}
\end{figure}

However, the situation changes for larger values of $\alpha_r$. As
shown in Fig.~\ref{fig2}, for higher values of
$\alpha_r$ these maxima and minima disappear and the difference
(distance) between the finite temperature and $T=0$ curves increase.
In other words, increasing ($\alpha_r$) leads to the suppression
of the extremum points in the $E(R)$-curve.

\begin{figure}[htbp]
\begin{center}
\includegraphics[width=9cm, height=6.5cm]{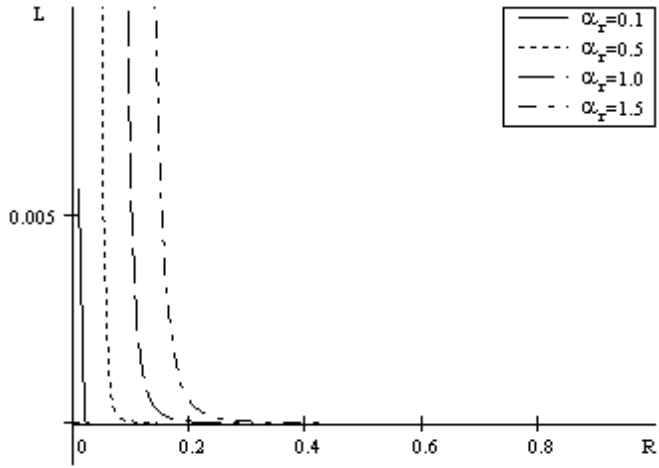}
\end{center}
\caption{The quantity $L(R)$ for different values of $\alpha_r
=0.1$.} \label{fig3}
\end{figure}

\begin{figure}[htbp]
\begin{center}
\includegraphics[width=9cm, height=6.5cm]{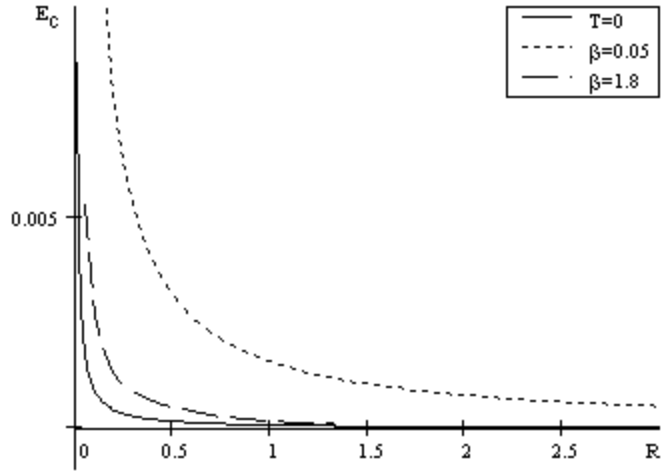}
\end{center}
\caption{Casimir energy as a function of $R$ for different
temperatures at $\alpha_r =0$.} \label{fig4}
\end{figure}

\begin{figure}[htbp]
\begin{center}
\includegraphics[width=15cm, height=6cm]{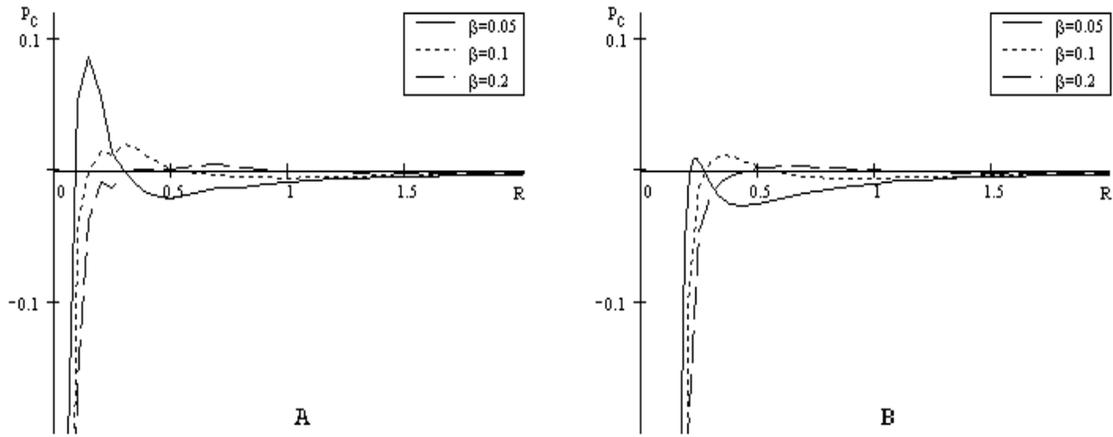}
\end{center}
\caption{Dependence of the Casimir pressure on the sphere radius
for different temperatures and fixed $\alpha_r$.
 A: for $\alpha_r =10^{-6}$; B: $\alpha_r =0.1$.} \label{fig5}
\end{figure}

This can be understood if results in Fig.~\ref{fig3}, where
the integral $L$ is plotted as a function of $R$ for different
values of $\alpha_r$ are considered. It is clear in this plot that by
increasing $\alpha_r$ the curve shifts from right to left side
i.e. the compactification parameter is higher as the $L(R)$-curve is closer
to the vertical axis. As the only difference in $E(R)$ for
different values of $\alpha_r$ arises from the integral
$L(R,\alpha_r)$ the difference between various $E(R)$
corresponding to different values of $\alpha_r$ is similar to that
for $L(R)$ at various $\alpha_r$.
The comparison of the $R$-dependence of Casimir energy for $\beta
=1.2$, $\beta =1.5$ ($\alpha_r =0$) and $T=0$ (from~\ci{bender}) in Fig.~\ref{fig4} shows
that our results approach those obtained
by Bender and Hays, as temperature
approaches zero. In Fig.~\ref{fig5} plots for the Casimir pressure
for various values of $\beta$ are presented. Three values of
temperature are considered in this case, $\beta =0.2$, $\beta
=0.1$ and $\beta =0.05$ for $\alpha_r =10^{-6}$ and $\alpha_r
=0.1$. It is clear from these plots that at certain values of $R$
corresponding to maxima and minima of the curve $E(R)$ the
function $P(R)$ changes its sign one or two times depending on
the value of $\beta$.
Therefore one may consider the critical temperature where
the Casimir pressure becomes negative. This critical temperature
depends on the radius of the sphere, i.e. for different values of $R$
we get different values of the critical temperature. In addition, it
depends on the value of the compactification parameter, $\alpha_r$ whose
increase leads to making a "smooth" $E(R)$ curve and the maxima and
minima in this curve are suppressed. In other words, in the
$R$-dependence of the Casimir energy, the increase of $\alpha_r$ plays
a similar role as that of increasing $\beta$, i.e., decreasing of
$T$.

\section{Conclusions}
Thus we have treated Casimir effect for spherical boundaries at finite temperature using the prescription
of generalized thermofield dynamics. Within this approach the heat bath effects and space compactification
effects (which means the presence of the spatial periodicity) in the zero-point energy and Casimir pressure are studied.
The results show that the dependence of the Casimir energy on the spherical radius is considerably different than that in the case of zero temperature results obtained by Bender and Hays and others. The first difference is the fact
that, unlike the case of the zero temperature, the $R$-dependence of the
Casimir energy is not factorized in the case of $T\not= 0$. This leads to other differences between
finite and zero temperature results. In particular as temperature decreases the form of $E(R)$-curve changes and for higher values of temperature maxima and minima can appear in this curve. This implies a change of the sign for the Casimir pressure. However,
such changes can be suppressed by increasing ($\alpha_r$) and, in addition, these extremum points can be suppressed and curve becomes smooth.
We note that the heat bath effects become considerable for rather small values of $\beta$. It is clear from Fig.~\ref{fig3} that the difference between zero-temperature and finite temperature Casimir energies disappears for values of $\beta$ higher than 1.5. Therefore the "convenient" candidate system where finite temperature effects can be found should be considered among systems
coupled to a strongly interacting environment, such as hadrons in quark-gluon plasma. This would imply that the magnitude of the de-confining temperature may be affected by the Casimir energy. Finally it is important to emphasize that the use of generalised
Bogoliubov transformations provides a different perspective to the presence of Casimir energy i.e. it is a manifestation of condensation of the scalar field in the vacuum. It may be anticipated that for a real nucleon with quarks and gluons as its content in a spherical surroundings, the Casimir energy may be viewed as a condensation of quark and gluon fields in the vacuum. These considerations make a study of a real nucleon with QCD field involving quarks and gluons a rather fascinating subject to explore. Such a study is presently in progress.

\section*{Acknowledgements}

FCK thanks NSERCC for financial support. AES thanks
CNPQ(Brazil) for financial assistance. The work of DUM and Kh.Yu.R is supported by the grant of the Uzbek Academy of Sciences (FA-F2-F084). The work of Kh.T.B is supported by the INTAS Fellowship (No. 06-1000014-6418).

\end{document}